\documentstyle[12pt,fleqn]{article}
\newcommand{\cZ}{{\cal{Z}}}
\newcommand{\cC}{{\cal{C}}}
\newcommand{\Z}{{Z \!\!\! Z}}
\newcommand{\cD}{{\cal{D}}}
\newcommand{\beq}{\begin{equation}}
\newcommand{\eeq}{\end{equation}}
\newcommand{\beqn}{\begin{eqnarray}}
\newcommand{\eeqn}{\end{eqnarray}}
\newcommand{\CK}[1]{\mbox{\scriptsize c}_{\mbox{$\scriptstyle #1$}}}
\newcommand{\nsum}[2]{\sum_{ #1(\CK{#2}) \in \Z }}

\newcommand{\nddsum}[2]{\sum_{\stackrel{\scriptstyle \dual #1(\dual\CK{#2})
\in \Z} {\delta \dual #1=0}}}
\newcommand{\dd}{\mbox{d}}
\newcommand{\dual}{\mbox{}^{\ast}}
\newcommand{\LL}{{I\!\! L}}
\newcommand{\intpi}{\int\limits_{-\pi}^{+\pi} {\cD}}
\newcommand{\intinf}{\int\limits_{-\infty}^{+\infty} {\cD}}
\newcommand{\expb}[1]{\exp\left\{ #1 \right\} }
\newcommand{\ie}{\hbox{\it i.e.}{}}
\newcommand{\etc}{\hbox{\it etc.}{}}
\newcommand{\eg}{\hbox{\it e.g.}{}}
\newcommand{\re}[1]{(\ref{#1})}
\newcommand{\half}{\frac 12}
\newcommand\Appendix[1]{\par
\setcounter{section}{0}
 \setcounter{equation}{0}
 \renewcommand{\thesection}{Appendix \Alph{section}}
\section{#1}
 \def\theequation{\Alph{section}.\arabic{equation}}}

\def\JP{ J.~Phys.}
\def\NP{ Nucl.~Phys.}
\def\ZP{ Z.~Phys.}
\def\PL{ Phys.~Lett.}
\def\PRL{ Phys.~Rev.~Lett.}

\newcommand{\AmS}{{\protect\the\textfont2
  A\kern-.1667em\lower.5ex\hbox{M}\kern-.125emS}}

\title{
Various Abelian Projections of $SU(2)$ Lattice Gluodynamics and
Aharonov-Bohm Effect in the Field Theory \thanks{This material is based on
the talks on the Lattice93 symposium (Dallas October 1993)
by M.Zubkov "Aharonov-Bohm effect in lattice gluodynamics" and by
M.Polikarpov "String representation of the Abelian Higgs theory."}}
\author{M.N.Chernodub, M.I.Polikarpov, M.A.Zubkov}

\date{}

\begin{document}

\begin{titlepage}
\maketitle

\mbox{}

\thispagestyle{empty}
\mbox{}
\begin{abstract}

        We show that in general abelian projection of lattice
gluodynamics it is not only monopoles but also strings are present. Both
these topological excitations may be responsible for the confinement of
color. We illustrate our ideas by some explicit results in the Abelian Higgs
model with the Villain action.

\end{abstract}

\end{titlepage}

\section{Introduction}

The idea of the abelian projection \cite{tHoo81} is to reduce the
gluodynamics to an abelian system, in which we can understand all physical
effects including confinement. The last year's review by Suzuki
\cite{Suz92} and many other talks given at the present symposium (sections
``QCD vacuum'') contain various numerical and analytical results confirming
the monopole confinement mechanism in $SU(2)$ lattice gluodynamics. All
these results are obtained for the so-called maximal abelian projection
\cite{KrScWi87,KrLaScWi87}, in which the gauge transformation makes the link
matrices diagonal "as much as possible". Formally, the matrices of the
gauge transformation $\Omega_x$ maximize the quantity

\beq
        R(U') = \sum_{x,\mu} Tr(U'_{x\mu}\sigma_3 U'^{+}_{x\mu}\sigma_3)\;\;,
                        \label{R}
\eeq
\beq
        U'_{x\mu} = \Omega^+_x U_{x\mu} \Omega_{x+\hat{\mu}}\;\;.
\eeq

        The other abelian projections (such as the diagonalization of the
plaquette matrix $U_{x,12}$) do not give evidence that vacuum behaves as
the dual superconductor. Below we give two examples. First, it occurs
\cite{IvPoPo90} that the fractal dimensionality of the monopole currents
extracted from the lattice vacuum by means of the maximal abelian projection
is strongly correlated with the string tension. If monopoles are extracted
by means of the other projections (diagonalization of $U_{x,12}$) this
correlation is absent \cite{IvPoPo93}. The other example is the temperature
dependence of the monopole condensate $C$ measured on the basis of the
percolation properties of the clusters of the monopole currents
\cite{IvPoPo93}. For the maximal abelian projection the condensate is
non-zero below the critical temperature $T_c$ and vanishes above it. For the
projection which corresponds to the diagonalization of $U_{x,12}$, the
condensate is nonzero at $T>T_c$. The last result was obtained by the
authors of \cite{IvPoPo93}, but has not been published.

        In Sect.~2 we discuss why the maximal abelian projection may be the
unique projection in which only the monopoles are responsible for the
confinement. In Sect.~3 we argue that for a general projection not only
monopoles but also strings are present in the abelian vacuum. In Sect.~4 we
discuss a simplified abelian model in which confinement of quarks is due to
the interaction with strings. It occurs that this interaction is the field
theoretical analogue of the Aharonov-Bohm effect.

\section{Confinement in the Maximal Abelian Gauge.}

        There is a small parameter $\varepsilon$ in lattice
gluodynamics in the maximal abelian gauge. This parameter is the
natural measure of closeness between the diagonal matrices and the link
matrices after the gauge projection.  For the $SU(2)$ gluodynamics, we have

\beq
        \varepsilon^2=\frac{<\sum_{x,\mu}\mid
        U^{12}_{x\mu}\mid^2>}{<\sum_{x,\mu}\mid U^{11}_{x\mu}\mid^2>} \;\; ,
        \label{eps}
\eeq
where $U^{ij}_{x\mu}$ are the matrix elements of the link matrix
$U_{x\mu}$.

        If we neglect the fluctuations of $\mid U^{11}_{x\mu} \mid$, as well
as the gauge fixing term and the Faddeev-Popov determinant, the
$SU(2)$ action in the minimal abelian gauge becomes:

\beq
        S_{P}=\beta\frac{1}{2} Tr U_{P} =\bar{\beta}\cos\theta_{P} +
        O(\varepsilon^2) +  ... \;\; ,
\eeq

where $\bar{\beta}=\beta\mid U_{11} \mid ^4$, $\theta_P$ is the plaquette
angle constructed in the usual way from the link angles $\theta_{x\mu}
\; , U^{11}_{x\mu}=\mid U^{11}_{x\mu}\mid {e^{i\theta_{x\mu}}}$.

        Thus, in the naive approximation, we get the action of the compact
electrodynamics.  Since in the compact electrodynamics the confinement is
due to the monopole condensation, it is not surprising that, in the maximal
abelian gauge, the vacuum of gluodynamics is an analogue of the dual
superconductor.

        Of course, this is only an intuitive argument. The confinement in
the $U(1)$ theory exists in the strong coupling region. Therefore, in order
to explain the confinement at large values of $\beta$ in $SU(2)$
gluodynamics, we have to study in detail the special features of the gauge
fixing procedure (such as Faddeev-Popov--determinant, gauge fixing term,
fluctuations of $\mid U^{11}_{x\mu}\mid$ etc.).

\section{Topological Excitations in the General Abelian Projection.}

        Now we consider the general abelian projection of the $SU(2)$ lattice
gluodynamics. In this case there is no small parameter, similar to
$\varepsilon$ \ref{eps}, and not only diagonal elements $U^{11}_{x\mu}$ but
also nondiagonal elements $U^{12}_{x\mu}$ contributes to the effective
action.  It is convenient to consider the following parametrization of the
link matrix:

\beq
                U_{x\mu}=\exp\{\frac{i}{2}(\rho^1_{x\mu}\sigma_1 +
        \rho^2_{x\mu}\sigma_2)\}\exp\{\frac{i}{2}\theta_{x\mu}\sigma_3\}.
\eeq
After the abelian projection, the fields $\theta_{x\mu}$ and
$\rho_{x\mu}=\rho^1_{x\mu}+i\rho^2_{x\mu}$ transform under the abelian gauge
transformation as follows:

\beq
        \theta_{x\mu}\to \theta_{x\mu} +\alpha_x -\alpha_{x+\hat{\mu}}
         \nonumber
\eeq

\beq
        \rho_{x\mu}\to e^{2i\alpha_x}\rho_{x\mu} \;\; ,
\eeq
so the diagonal gluons become the gauge field (photon) $\theta$,
and the non-diagonal gluons play the role of the charge 2 matter vector
field $\rho$, \cite{tHoo81}. For the general abelian projection we have

\beq
<\sum_{\mu}\mid\rho_{x\mu}\mid^2 >\neq 0\;\;,
\eeq
therefore we have the condensed
matter field and in this system the counterpart of the Nielsen--Olesen
strings exists. These topological excitations, as shown later, may play as
important role in the confinement mechanism as the monopoles in the maximal
abelian projection. Using standard algorithm, we can extract the
monopoles from a given configuration of the compact abelian gauge fields on
the lattice.  The Nielsen--Olesen strings can be extracted in a similar way
from a given configuration of the compact matter field and gauge field.
Consider, for simplicity, a condensed scalar matter field, for
instance, the composite field

\beq
    \Phi_x = \sum_{\mu}\rho_{x\mu} = \mid \Phi_x\mid e^{i\varphi_x} \;\;.
\eeq

        Below we use the notations of the calculus of differential forms on
the lattice, which are briefly described in Appendix A. Consider the
gauge invariant link variable $L=d\varphi +2\theta +2\pi l$ and the
plaquette variable $P=d\theta +2\pi n$, where $l$ and $n$ are integer
numbers such that $-\pi <L\leq\pi$, and $-\pi <P\leq\pi$. The world sheets
$\sigma$ of the Nielsen--Olesen strings and world trajectories $j$ of the
monopoles are defined on the dual lattice by:

\beq
        \sigma = \frac{1}{2\pi}(\dual \dd L-2 \dual P) = \dual \dd l -2
        \dual n \;\;,
\eeq

\beq
        j =\frac{1}{2\pi} \dual \dd P = \dual \dd n \;\; .
\eeq
Since $\delta\sigma = j$, the strings may be opened, and may carry
monopole and antimonopole at the ends; the monopole currents form closed
lines:  $\delta j = 0$.
Thus, in general abelian projection gluodynamics is reduced to an
abelian Higgs model in which the topological excitations are monopoles and
strings.

\section{Abelian Higgs Theory and Aharonov--Bohm Effect in the Field Theory}

We consider now the Abelian Higgs model with noncompact gauge field, in
order to study the contribution of the strings into the confinement
mechanism. The contribution of monopoles, which exist due to the compactness
of the gauge field, is widely discussed in the literature. The
partition function of the $2D$ $XY$-model is equivalent to the partition
function of the Coulomb gas \cite{Ber70,KoTh73}, the partition function of
the $4D$ compact electrodynamics can be represented as a sum over the
monopole--antimonopole world lines \cite{BaMeKo77}. Now
we show that the partition function of the four--dimensional
Abelian Higgs theory can be represented as a sum over closed surfaces
which are the world sheets of the Abrikosov--Nielsen--Olesen strings
\cite{Abr57,NiOl73}.

        We consider the model describing interaction of the
noncompact gauge field $A_\mu$ with the scalar field $\Phi =
|\Phi|e^{i\varphi}$, whose charge is $Ne$ (to be in agreement with the
previous section, we have to set $N=2$, since the charge of the nondiagonal
gluon is $2e$). The selfinteraction of the scalar field is described by the
potential $V(\Phi) = \lambda (|\Phi|^2 - \eta^2)^2$. For simplicity, we
consider the limit as $\lambda \rightarrow \infty$, so that the radial part
of the scalar field is frozen, and the dynamical variable is compact:
$\varphi \in (-\pi,\pi]$.  The partition function for the Villain form of
the action is given by:

\beqn
 \cZ = \intinf A \intpi \varphi \nsum{l}{1}\expb{ -S_l(A,\dd\varphi)},
                                                \label{ZNC}
\eeqn
where
\beq S_l(A,\dd \varphi) = \frac{1}{2e^2}\|\dd A\|^2 +
        \frac{\kappa}{2} \| \dd\varphi + 2\pi l - N A \|^2 .  \label{ZNCS}
\eeq

We use the notations of the calculus of differential forms on the lattice
described in Appendix A. The symbol
$\int\cD\varphi$ ($\int\cD A$) denotes the integral over all site (link)
variables $\varphi$ ($A$). Fixing the gauge $\dd\varphi = 0$, we get the
following expression for the action \re{ZNCS}:  $S_l =
\frac{1}{2e^2}[A,(\delta\dd + N^2 \kappa e^2) A] + \ (terms \ linear \ in \
A)$; therefore, due to the Higgs mechanism, the gauge field acquires the
mass $m = N \kappa^\half e$; there are also soliton sectors of the Hilbert
space which contain Abrikosov--Nielsen--Olesen strings hidden in
the summation variable $l$ in \re{ZNC}.

        The partition function of the compact electrodynamics can be
represented as a sum over closed world lines of monopoles
\cite{BaMeKo77}. In the same way the partition function \re{ZNC} can be
rewritten as a sum over closed world sheets of the Nielsen--Olesen
strings\footnote{We use the superscript BKT,
since a similar transformation of the partition function was first found
by Berezinskii \cite{Ber70}, Kostrlitz and Thouless \cite{KoTh73}, who
showed that the $XY$ model is equivalent to the Coulomb gas in two
dimensions.}:

\beqn
\cZ ^{BKT}= const. \nddsum{\sigma}{2} \expb{- 2\pi^2\kappa (\dual
\sigma,(\Delta + m^2)^{-1}\dual \sigma)}. \label{ZNCBKT}
\eeqn
A derivation of this representation is given in \cite{PoWiZu93}.
The sum here is over the integer variables $\dual \sigma$
attached to the plaquettes $\dual c_2$ of the dual lattice. The condition
$\delta \dual \sigma = 0$ means that for each link of the dual lattice the
``conservation law'' is satisfied: $\sigma_1 + \sigma_2 + \sigma_3 =
\sigma_4 + \sigma_5 + \sigma_6$, where $\sigma_i$ are integers
corresponding to plaquettes connected to the considered link. The signs of
$\sigma_i$'s in this ``conservation law'' are dictated by the definition of
$\delta$ (by the orientation of the plaquettes 1,...,6). If $\sigma=0,1$,
then the condition $\delta \dual \sigma = 0$ means that we consider closed
surfaces made of plaquettes with $\sigma = 1$. In \re{ZNCBKT} we have
$\sigma \in \Z$, which means that a single plaquette may ``decay'' into
several ones, but still the surfaces made of plaquettes with $\sigma \neq
0$ are closed. It follows from \re{ZNCBKT} that the strings interact with
each other via the Yukawa forces\footnote{Due to the definition of the
integration by parts $(\varphi,\delta\psi) = (\dd\varphi,\psi)$, the
operator \mbox{$(\Delta + m^2)^{-1}$} (and not \mbox{$(-\Delta +
m^2)^{-1}$}) is positive definite on the Euclidean lattice.},
$(\Delta +m^2)^{-1}$.

        Now we calculate the quantum average of the Wilson loop for the
charge $Me$, $W_M(\cC) = \expb{i M (A,j_\cC)}$, in the BKT representation.

Repeating all steps which transform \re{ZNC} into \re{ZNCBKT}, we get:

\beqn
 <W_M(\cC)> =  \frac{1}{\cZ^{BKT}} \nonumber \\
 \times\nddsum{\sigma}{2} \exp\{-
2\pi^2\kappa (\dual\sigma,(\Delta + m^2)^{-1}\dual \sigma)
 - \frac{M^2 e^2}{2}( j_\cC,(\Delta + m^2)^{-1} j_\cC) \nonumber \\
 - 2 \pi i \frac{M}{N} (j_\cC,(\Delta + m^2)^{-1}\delta \sigma)
 + 2 \pi i \frac{M}{N} \LL(\sigma,j_\cC) \}\;\;.  \label{WNC}
\eeqn
The first three terms in the exponent describe the short--range (Yukawa)
interactions: surface -- surface, current -- current and current -- surface.
In spite of the gauge field acquiring the mass $m =N \kappa^\half e$, there
a is {\it long--range} interaction of geometrical nature, described
by the last term in the exponent: $\LL (\sigma,j_\cC)$, $\LL$ being the
four--dimensional analogue of the Gauss linking number for loops in three
dimensions, \ie\ the linking number of surfaces defined by
$\{\sigma\}$ and loop defined by $j_\cC$. The explicit expression for $\LL$
is:

\beq
       \LL = (\dual j_\cC, {\Delta}^{-1} \dd \dual \sigma) =
       (\dual j_\cC, \dual n)\;\;,              \label{L}
\eeq
where $\dual n$ is an integer valued 3-form which is the solution of the
equation $\delta \dual n = \dual \sigma$. It is clear now that
$\LL$ is equal to
the number of points at which the loop $j_\cC$ intersects the
three--dimensional volume $\dual n$ bounded by the closed surface defined by
$\dual \sigma(\dual c_2)$. The elements of the surface $\dual \sigma$ may
carry any integer number, so that any intersection point may contribute an
integer into $\LL$.
Therefore, $\LL$ is the linking number of
the world sheet of the strings and the current $j_\cC$ which define Wilson's
loop $W_M(\cC)$. The reason for the topological interaction is that the
charges $e,\ 2e, \ \ldots (N-1)e$ cannot be completely screened by the
condensate of the field of charge $Ne$; if $M/N$ is integer, then the
screening is complete and there is no topological interaction. From the
another point of view this, is the four--dimensional analogue
\cite{AlWi89,AlMaWi90,PrKr90} of the Aharonov--Bohm effect: strings
correspond to solenoids which scatter charged particles.

To apply the model in question to abelian projected gluodynamics, we simply
have to set $M=1,\ N=2$, since the charge of ``test infinitely heavy'' quark
is unity and the charge of the condensed nondiagonal gluon, whose phase plays
the role of the field $\varphi$, is two. The topological interaction can
yield the area low; this fact can be easily proven for completely random
surfaces $\sigma$.

Thus,we conclude that the confinement in gluodynamics can be due not only
to monopoles but also to strings.

MIP is grateful to Ken Yee and Dick Haymaker for interesting
discussions.
The work of MIP and MAZ has been partially supported by the
grant of the Russian Foundation for the Fundamental
Sciences. The work of MNC, MIP and MAZ has been partially supported by the
Japan Society for Promotion of Science (JSPS) Program on Japan-FSU
scientists Collaboration. The members of the delegation of Russia at the
Lattice 93 symposium expresses their thanks to International Science
Foundation and to the Organizing Committee for the financial support,
without which their participation in the symposium would be impossible.

\Appendix{ }

Here we briefly summarize the main notions from the theory of
differential forms on the lattice \cite{BeJo82}.  The advantages of the
calculus of differential forms consists in the general character of the
expressions obtained. Most of the transformations depend neither on the
space--time dimension, nor on the rank of the fields. With minor
modifications, the transformations are valid for lattices of any form
(triangular, hypercubic, random, \etc). A differential form of rank $k$ on
the lattice is a function $\phi_{k}$ defined on $k$-dimensional cells $c_k$
of the lattice, \eg\ the scalar (gauge) field is a 0--form (1--form). The
exterior differential operator {\it d} is defined as follows:

\beq
(\dd \phi ) (c_{k+1}) =\sum_{\CK{k} \in \partial\CK{k+1}} \phi(c_{k}).
\label{def-dd}
\eeq
Here $\partial c_{k}$ is the oriented boundary of the $k$-cell
$c_{k}$. Thus the operator {\it d} increases the rank of the form by unity;
$\dd \varphi$ is the link
variable constructed, as usual, in terms of the site angles $\varphi$, and
$\dd A$ is the plaquette variable constructed from the link variables $A$.
The scalar product is defined in the standard way:
if $\varphi$ and $\psi$ are $k$-forms, then
$(\varphi,\psi)=\sum_{c_k}\varphi(c_k)\psi(c_k)$, where $\sum_{c_k}$ is the
sum over all cells $c_k$.
To any $k$--form on the $D$--dimensional lattice there
corresponds a $(D-k)$--form $\dual\Phi(\dual c_k)$ on the dual lattice,
$\dual c_k$ being the $(D-k)$--dimensional cell on the dual lattice. The
codifferential $\delta=\dual \dd \dual$ satisfies the partial
integration rule: $(\varphi,\delta\psi)=(\dd\varphi,\psi)$.
Note that $\delta \Phi(c_k)$ is a $(k-1)$--form and
$\delta \Phi(c_0) = 0$. The norm is defined by: $\|a\|^2=(a,a)$; therefore,
$\|\dd\varphi+2\pi l\|^2$ in \re{ZNCS} implies summation over all links.
$\nsum{l}{1}$ denotes the sum over all configurations of the integers $l$
attached to the links $c_1$. The action \re{ZNCS} is invariant under the
gauge transformations $A' = A + \dd \alpha$, $\varphi' = \varphi + \alpha$
due to the well known property $\dd^2 = \delta^2 = 0$. The
lattice Laplacian is defined by: $\Delta = \dd\delta + \delta\dd$.

\end{document}